\documentclass[paper]{article}
\usepackage[dvips]{epsfig}
\usepackage{latexsym}
\usepackage{amsmath}
\usepackage{graphicx}

\def\spur#1{\mathord{\not\mathrel{{\mathrel{#1}}}}}

\begin{document}
\baselineskip 16pt
 \oddsidemargin=0pt

\begin{center}
\bigskip \makeatletter

\makeatother \makeatletter

{\LARGE  Radiative corrections to the lightest  KK states  in the $T^2/(Z_2\times Z_2')$ orbifold}{\LARGE  \ \vspace{15pt}%
\\[0pt]
}
\textbf{A.T. Azatov\footnote{Email: aazatov@umd.edu}}\\[5pt]
\vspace{0pt}\textit{Department of Physics, University of Maryland,
College Park, MD 20742, USA}\vspace{0pt}\\[0pt]

\bigskip

\bigskip

\bigskip

\bigskip

\bigskip \bigskip \vspace{55pt} \texttt{Abstract}
\end{center}

\bigskip
We study radiative corrections localized at the fixed points of
the orbifold for the field theory in six dimensions with two
dimensions compactified on the $T_2/(Z_2\times Z_2')$ orbifold in
a specific realistic model for low energy physics that solves the
proton decay and neutrino mass problem. We calculate corrections
to the masses of the lightest stable KK modes, which could be the
candidates for  dark matter.

\newpage
\section{Introduction}
One of the important questions in particle physics today is the
nature of physics beyond the standard model (SM). The new Large
Hadron Collider (LHC) machine starting soon, experiments searching
for dark matter of the universe as well as many neutrino
experiments planned or under way, have raised the level of
excitement in the field since they are poised to provide a unique
experimental window into this new physics.  The theoretical ideas
they are likely to test are supersymmetry, left-right symmetry as
well as possible hidden extra dimensions
\cite{antoniadis}\cite{dvali}\cite{kokorelis} in nature, which all
have separate motivations and address different puzzles of the SM.
In this paper, I will focus on an aspect of one interesting class
of models known as universal extra dimension models(UED)
\cite{acd}(see for review \cite{review}). These models provide a
very different class of new physics at TeV (see \cite{higgsboson}
for the constraints on size of compactification $R $ ) scale than
supersymmetry. But in general UED models based on the standard
model gauge group, there is no simple explanation for the
suppressed proton decay and the small neutrino mass. One way to
solve the proton decay problem in the context of total six
space-time dimensions, was proposed in \cite{protonstability}. In
this case, the additional dimensions lead to the new U(1)
symmetry, that suppresses all baryon-number violating operators.
The small neutrino masses can be explained by the propagation of
the neutrino in the seventh warp extra dimension
\cite{neutrinomass}. On the other hand we can solve both these
problems by extending gauge group to the $SU(2)_L\times
SU(2)_R\times U(1)_{B-L}$\cite{lrs}. Such class of UED models were
proposed in \cite{mohapat}. In this case, the neutrino mass is
suppressed due to the $B-L$ gauge symmetry  and specific
orbifolding conditions that keep left-handed neutrinos at zero
mode and forbid lower dimensional operators that can lead to the
unsuppressed neutrino mass.

 An important consequence of UED models is the existence of a new class
of dark matter particle, i.e. the lightest KK (Kaluza-Klein)
mode\cite{tait}. The detailed nature of the dark matter and its
consequences for new physics is  quite model-dependent. It was
shown  recently \cite{ken} that the lightest stable KK modes in
the model \cite{mohapat} with universal extra dimensions could
provide the required amount of cold dark matter \cite{feng}. Dark
matter in this particular class of UED models is an admixture of
the KK photon and right-handed neutrinos. In the case of the two
extra dimensions, KK mode of the every gauge boson is accompanied
by the additional adjoint scalar which has the same quantum
numbers as a gauge boson. In the tree level approximation KK
masses of this adjoint scalar and gauge boson are the same, so
they both can be  dark matter candidates. The paper \cite{ken}
presented relic density analysis assuming either the adjoint
scalar or the gauge boson is the lightest stable KK particle.
These assumptions lead to the different restrictions on the
parameter space. My goal in this work was to find out whether
radiative corrections could produce mass splitting of these modes,
and if they do, determine the lightest stable one. In this
calculations I will follow the works \cite{Georgi},\cite{matchev}.
Similar calculations for different types of orbifolding were
considered in \cite{ponton} ($T^2/Z_4$ orbifold),and
\cite{quiros}($M_4\times S^1/Z_2$ and $T^2/Z_2$ orbifolds).

\section{Model}
In this sections we will review the basic features of the model
\cite{mohapat}. The gauge group of the model is $SU(3)_c\times
SU(2)_L\times SU(2)_R\times U(1)_{B-L}$ with the following matter
content for generation :
\begin{eqnarray}
{\cal Q}_{1,-}, {\cal Q}'_{1,-}= (3,2,1,\tfrac{1}{3});&
{\cal Q}_{2,+}, {\cal Q}'_{2,+}= (3,1,2,\tfrac{1}{3});\nonumber\\
{\cal \psi}_{1,-}, {\cal \psi}'_{1,-}= (1,2,1,-1);& {\cal
\psi}_{2,+}, {\cal \psi}'_{2,+}= (1,1,2,-1); \label{matter}
\end{eqnarray}

We denote the gauge bosons as $G_A$, $W^{\pm}_{L,A}$,
$W^{\pm}_{R,A}$, and $B_A$, for $SU(3)_c$, $SU(2)_L$, $SU(2)_R$
and $U(1)_{B-L}$ respectively, where $A=0,1,2,3,4,5$ denotes the
six space-time indices.  We will also use the following short hand
notations: Greek letters $\mu,\nu,\dots=0,1,2,3$ for  usual four
dimensions indices and lower case Latin letters $a,b,\dots=4,5$
for the extra space dimensions.

 We compactify the extra $x_4$, $x_5$ dimensions into a torus,
$T^2$, with equal radii, $R$, by imposing periodicity conditions,
$\varphi(x_4,x_5) = \varphi(x_4+ 2\pi R,x_5) = \varphi(x_4,x_5+
2\pi R)$ for any field $\varphi$. We impose the further
orbifolding conditions i.e. $ Z_2:\vec{y} \rightarrow -\vec {y}
~\mbox{and}~
 Z'_2: \vec{y}~'\rightarrow -\vec{y}~'$
where $\vec y = (x_4,x_5)$,  $\vec{y}~' = \vec{y} - (\pi R /2, \pi
R/2)$. The $Z_2$ fixed points will be located at the coordinates
$(0,0)$ and $(\pi R,\pi R )$, whereas those of $Z_2'$ will be in
$(\pi R/2,\pm \pi R/2)$. The generic field $\phi(x_\mu, x_a)$ with
fixed $Z_2\times Z_2' $ parities can be expanded as:
\begin{eqnarray}
\phi(+,+)=\frac{1}{2\pi R}\varphi^{(0,0)}+\frac{1}{\sqrt{2}\pi
R}\sum_{n_4+n_5~ \text{-even}}\varphi^{(n_4,n_5)}(x_\mu)
cos\left(\frac{n_4x_4 +n_5x_5
}{R}\right)\nonumber\\
\phi(+,-)=\frac{1}{\sqrt{2}\pi R}\sum_{n_4+n_5~\text{
-odd}}\varphi^{(n_4,n_5)}(x_\mu) cos\left(\frac{n_4x_4 +n_5x_5
}{R}\right)\nonumber\\
\phi(-,+)=\frac{1}{\sqrt{2}\pi R}\sum_{n_4+n_5~\text{
-odd}}\varphi^{(n_4,n_5)}(x_\mu) sin\left(\frac{n_4x_4 +n_5x_5
}{R}\right)\nonumber\\
\phi(-,-)=\frac{1}{\sqrt{2}\pi R}\sum_{n_4+n_5~\text{
-even}}\varphi^{(n_4,n_5)}(x_\mu) sin\left(\frac{n_4x_4 +n_5x_5
}{R}\right) \label{decompos}
\end{eqnarray}
One can see that only  the $(+,+)$ fields will have zero modes. In
the effective 4D theory the mass of each mode has the form:
$m_{N}^2 = m_0^2 + \frac{N}{R^2}$; with $N=\vec{n}^2=n_4^2 +
n_5^2$ and $m_0$ is the physical mass of the zero mode.

We assign the following $Z_2\times Z'_2$ charges to the various
fields:
\begin{eqnarray}
G_\mu(+,+);\quad B_\mu(+,+); \quad
W_{L,\mu}^{3,\pm}(+,+);W^3_{R,\mu}(+,+);
W^\pm_{R,\mu}(+,-); \nonumber\\
G_{a}(-,-);\quad B_a(-,-);\quad W_{L,a}^{3,\pm}(-,-);
W^3_{R,a}(-,-); W^\pm_{R,a}(-,+). \label{gparity}
\end{eqnarray}
As a result, the gauge symmetry $SU(3)_c\times SU(2)_L\times
SU(2)_R\times U(1)_{B-L}$ breaks down to $SU(3)_c\times
SU(2)_L\times U(1)_{I_{3R}}\times U(1)_{B-L}$ on the 3+1
dimensional brane. The $W^{\pm}_R$ picks up a mass $R^{-1}$,
whereas prior to symmetry breaking the rest of the gauge bosons
remain massless.

For quarks we choose,
\begin{eqnarray}
 Q_{1,L}\equiv
   \left(\begin{array}{c} u_{1L}(+,+)\\ d_{1L}(+,+)\end{array}\right);
 &\quad&
 Q'_{1,L}\equiv
   \left(\begin{array}{c} u'_{1L}(+,-)\\ d'_{1L}(+,-)\end{array}\right);
   \nonumber \\
 Q_{1,R}\equiv
   \left(\begin{array}{c} u_{1R}(-,-)\\ d_{1R}(-,-)\end{array}\right);
 &\quad&
 Q'_{1,R}\equiv
   \left(\begin{array}{c} u'_{1R}(-,+)\\ d'_{1R}(-,+)\end{array}\right);
   \nonumber
\end{eqnarray}
\begin{eqnarray}
 Q_{2,L}\equiv
   \left(\begin{array}{c} u_{2L}(-,-)\\ d_{2L}(-,+)\end{array}\right);
 &\quad&
Q'_{2,L}\equiv
   \left(\begin{array}{c} u'_{2L}(-,+)\\ d'_{2L}(-,-)\end{array}\right);
   \nonumber \\
 Q_{2,R}\equiv
   \left(\begin{array}{c} u_{2R}(+,+)\\ d_{2R}(+,-)\end{array}\right);
& \quad& Q'_{2,R}\equiv
   \left(\begin{array}{c} u'_{2R}(+,-)\\ d'_{2R}(+,+)\end{array}\right);
\label{quarks} \end{eqnarray} and for leptons:
 \begin{eqnarray}
 \psi_{1,L}\equiv
   \left(\begin{array}{c} \nu_{1L}(+,+)  \\ e_{1L}(+,+)\end{array}\right);
  &\qquad&
 \psi'_{1,L}\equiv
 \left(\begin{array}{c} \nu'_{1L}(-,+)  \\ e'_{1L}(-,+)\end{array}\right);
   \nonumber \\
 \psi_{1,R}\equiv
    \left(\begin{array}{c} \nu_{1R}(-,-)  \\ e_{1R}(-,-)\end{array}\right);
  & \qquad &
 \psi'_{1,R}\equiv
  \left(\begin{array}{c} \nu'_{1R}(+,-) \\ e'_{1R}(+,-)\end{array}\right);
   \qquad
   \nonumber \\ [1ex]
 \psi_{2,L}\equiv
   \left(\begin{array}{c} \nu_{2L}(-,+) \\ e_{2L}(-,-)\end{array}\right);
  &\qquad&
 \psi'_{2,L}\equiv
    \left(\begin{array}{c} \nu'_{2L}(+,+)\\ e'_{2L}(+,-)\end{array}\right);
  \nonumber \\
 \psi_{2,R}\equiv
 \left(\begin{array}{c} \nu_{2R}(+,-)\\ e_{2R}(+,+)\end{array}\right);
&   \qquad &
 \psi'_{2,R}\equiv
  \left(\begin{array}{c} \nu'_{2R}(-,-)\\ e'_{2R}(-,+)\end{array}\right).
\label{leptons}
 \end{eqnarray}
 The zero modes i.e. (+,+) fields correspond to the standard
 model fields along with an extra singlet neutrino which is
 left-handed. They will have zero mass prior to gauge symmetry
 breaking.

The Higgs sector of the model consists of
\begin{align}
\phi &\equiv
\left(\begin{array}{cc} \phi^0_u(+,+) & \phi^+_d(+,-)\\
   \phi^-_u(+,+) &  \phi^0_d(+,-)\end{array}\right);\nonumber\\
\chi_L&\equiv \left(\begin{array}{c} \chi^0_L(-,+) \\
   \chi^-_L(-,+)\end{array}\right); \quad
\chi_R\equiv \left(\begin{array}{c} \chi^0_R(+,+) \\
   \chi^-_R(+,-)\end{array}\right),
 \end{align}
with the charge assignment under the gauge group,
\begin{eqnarray}
\phi &=& (1,2,2,0),\nonumber\\
\chi_L&=&(1,2,1,-1),\quad \chi_R=(1,1,2,-1).
\end{eqnarray}
In the limit when the scale of $SU(2)_L$ is much smaller than the
scale of $SU(2)_R$ (that is, $v_w\ll v_R$) the symmetry breaking occurs in
two stages.
First $SU(2)_L\times SU(2)_R\times U(1)\rightarrow
SU(2)_L\times U(1)_Y$, where a linear combination of $B_{B-L}$ and
$W_{R}^3$, acquire a mass to become $Z'$, while orthogonal
combination of $B_{B-L}$ remains massless and serves as a
gauge boson for residual group $U(1)_Y$.  In terms of the
gauge bosons of $SU(2)_R$ and $U(1)_{B-L}$, we have
\begin{eqnarray}
Z'_A=\frac{g_RW_{R,A}^3-g_{B-L}B_{B-L,A}}{\sqrt{g_R^2+g_{B-L}^2}},\nonumber\\
B_{Y,A}=\frac{g_RB_{B-L,A}+g_{B-L}W_{R,A}^3}{\sqrt{g_R^2+g_{B-L}^2}}.
\end{eqnarray}
Then we have standard breaking of the electroweak symmetry. A
detailed discussion of the spectrum of the zeroth and first KK
modes was presented in \cite{ken}. The main result of the
discussion is that in the tree level approximation only the
 KK modes $B_{Y,\mu}$ $B_{Y,a}$ and $\nu_2$ will be
stable and can be considered as  candidates for dark matter, and
the relic CDM density value leads to the  upper limits on $R^{-1}$
of about 400-650 Gev, and the mass of the $M_{Z'}\leq 1.5$ Tev.
However, radiative corrections can split the KK masses of the
$B_{Y,\mu}$ and $B_{Y,a}$, and only the lightest of them will be
stable. The goal of this work is to find out which of the two
modes is lighter and serves as dark matter.

\section{Propagators}
To calculate the radiative corrections, we follow the methods
presented in Refs.~\cite{Georgi} and \cite{matchev}. We derive the
propagators for the scalar, fermion, and vector fields in the
$T^2/(Z_2\times Z_2')$ orbifold,  $\varepsilon$ and $\varepsilon'$
are the $Z_2$ and $Z_2'$ parities respectively, so arbitrary field
satisfying the boundary conditions
\begin{eqnarray}
\phi(x_4,x_5)=\varepsilon\phi(-x_4,-x_5)\nonumber\\
\phi(x_4,x_5)=\varepsilon'\phi(\pi R-x_4,\pi R-x_5) \label{orb}
\end{eqnarray}
can be decomposed as (we always omit the dependence on 4D coordinates)
\begin{align}
\phi(x_4,x_5)&=\Phi(x_4,x_5)+\varepsilon\Phi(-x_4,-x_5)\nonumber\\
&\quad+\varepsilon'\Phi(\pi
R-x_4,\pi R-x_5 )+\varepsilon\varepsilon '\Phi(x_4-\pi R,x_5-\pi
R).
\end{align}
The field $\phi$ will automatically  satisfy the orbifolding
conditions of Eq.~(\ref{orb}), and one can easily calculate
$\langle 0|\phi(x_4,x_5)\phi(x'_4,x'_5)|0\rangle$ in the momentum
space. This leads to the following expressions for the propagators
of the scalar, gauge and fermion fields. Propagator of the scalar
field is given by
\begin{eqnarray}
iD=\frac{i}{4(p^2-p_a^2)}\left(1+\varepsilon_\phi\varepsilon^{\prime}_\phi
e^{i p_a(\pi R)_a}\right) \left(
\delta_{p_ap^{\prime}_a}+\varepsilon_\phi\delta_{p_a
-p^{\prime}_a}\right),
\end{eqnarray}
where $ p_a(\pi R)_a\equiv\pi R( p_4+p_5)$. Propagator of the
gauge boson in($\xi=1$) gauge is
\begin{eqnarray}
iD_{AB}=\frac{-ig_{AB}}{4(p^2-p_a^2)}\left(1+\varepsilon_A\varepsilon_{A}
^{\prime} e^{i p_a(\pi R)_a}\right) \left(
\delta_{p_ap^{\prime}_a}+\varepsilon_A\delta_{p_a
-p^{\prime}_a}\right),
\end{eqnarray}
where fields $A_a$ and $A_\mu$ will  have opposite $Z_2\times
Z_2'$ parities: ($\varepsilon_\mu
=-\varepsilon_a,~\varepsilon'_\mu =-\varepsilon'_a$). The fermion
propagator is given by
\begin{eqnarray}
iS_{F}=\frac{i}{4(\mathord{\not\mathrel{{\mathrel{p}}}}-\mathord{\not%
\mathrel{{\mathrel{p_a'}}}})} \left( 1+\varepsilon _{\psi
}\varepsilon^{\prime}_{\psi }e^{ip_{a}(\pi R)_{a}}\right) \left(
\delta _{p_{a}p_{a}^{\prime }}+\Sigma_{45}\varepsilon _{\psi
}\delta _{p_{a}-p_{a}^{\prime }}\right),
\end{eqnarray}
where we have defined
\begin{eqnarray}
\spur{p_a}\equiv p_4\Gamma_4+p_5\Gamma_5.
\end{eqnarray}

\section{Radiative corrections to the fermion mass}
Now we want to find corrections for the mass of the $\nu_2$ field.
First let us consider general interaction between a fermion and a
vector boson,
\begin{eqnarray}
\mathcal{L}_{int}=g_{6D}\overline{\psi}\Gamma_A\psi A^A,
\end{eqnarray}
where $g_{6D}$ is 6 dimensional coupling
constant that is related to the 4 dimensional coupling $g$ by
\begin{eqnarray}
g=\frac{g_{6D}}{(2\pi R)}.
\end{eqnarray}
The gauge interaction will give mass corrections due to the
diagram Fig.1 (a).
\begin{figure}[h]
\includegraphics[scale=0.4,angle=90]{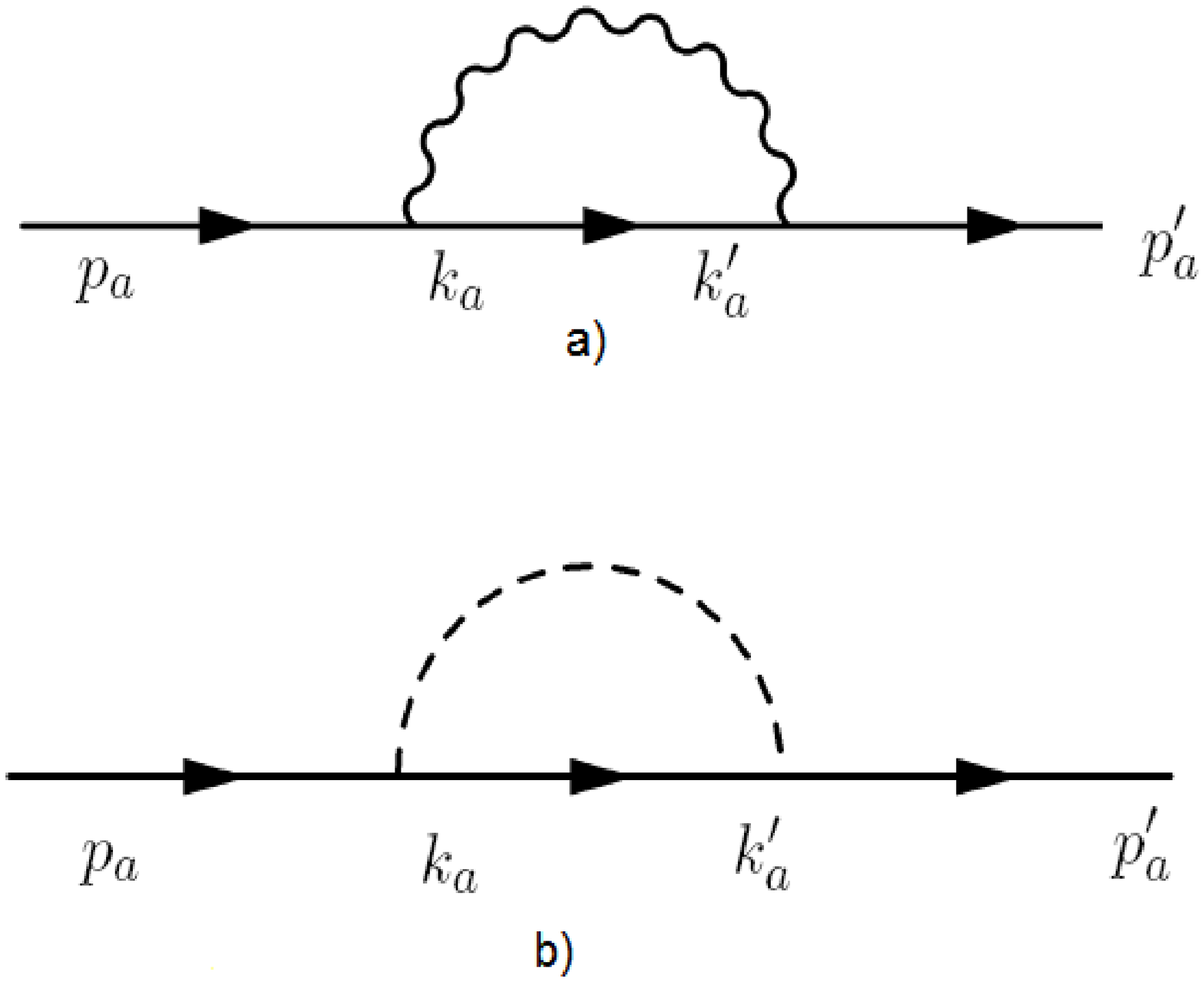} \label{figferm1} \caption{Fermion self
energy diagrams}
\end{figure}
The matrix element will be proportional to the
\begin{align}
i\Sigma &=-\sum_{k_{a}}g^{2}\int \frac{d^{4}k}{(2\pi )^{4}}\frac{1}{4}\frac{1%
}{(p-k)^{2}-(p_{a}-k_{a})^{2}}
\nonumber\\
&\quad\quad\times\Gamma_A\frac{\mathord{\not\mathrel{{%
\mathrel{k}}}}- \mathord{\not\mathrel{{\mathrel{k_a'}}}}}{k^{2}-k_{a}^{2}}[%
\delta_{k^{\prime}_a
k_a}+\varepsilon_\psi\Sigma_{45}\delta_{-k^{\prime}_a,k_a}]\Gamma^A[\delta_{(p-k)_a,(p^{\prime}-k^{\prime})_a}+\epsilon^A%
\delta_{(p-k)_a,-(p^{\prime}-k^{\prime})_a}]
\label{sigma}
\end{align}
The $\varepsilon_\psi$ and $\varepsilon'_\psi$, are the $Z_2\times
Z_2' $ parities of the fermion and $\varepsilon^A~,
{\varepsilon^A}'$ are the parities of the gauge boson. The sum is
only over the $k_{a}$ which are allowed by the $Z_{2} \times
Z_{2}'$ parities i.e. for the ones where $1+\varepsilon _{\psi
}\varepsilon _{\psi }^{\prime }e^{ik_{a}(\pi R)_{a}}\neq 0$
There are two types of terms that can lead to the corrections of
the fermion self energy, the bulk terms appearing due to the
nonlocal Lorentz breaking effects and brane like terms which
appear because of the specific orbifold conditions, but the bulk
terms for fermion self energy graph  appear to vanish (see
\cite{matchev}), so we will concentrate our attention only on the
brane like terms. In the case of our $T^2/(Z_2\times Z_2')$
orbifold they will be localized at the points $(0,0),~(\pi R,\pi R
),~(\pi R/2,\pm\pi R/2)$ (see Appendix). The numerator of the
integrand simplifies to
\begin{align}
\Gamma_A(\mathord{\not\mathrel{{\mathrel{k}}}}-\mathord{\not\mathrel{{%
\mathrel{k_a'}}}})\Gamma^A\varepsilon^A\delta_{(p+p^{\prime})_a,2k_a}+
\Gamma_A(\mathord{\not\mathrel{{\mathrel{k}}}}-\mathord{\not\mathrel{{%
\mathrel{k_a'}}}})\Sigma_{45}\Gamma^A\varepsilon_\psi\delta_{(p-p^{%
\prime})_a,2k_a}\nonumber\\
=4\mathord{\not\mathrel{{\mathrel{k_a'}}}}\varepsilon^\mu\delta_{p_a+p^{%
\prime}_a,2k_a}+4\mathord{\not\mathrel{{\mathrel{k_a'}}}}\varepsilon_\psi%
\Sigma_{45}\delta_{2k_a,p_a-p^{\prime}_a}
\end{align}
We can then write Eq.~(\ref{sigma}) as
\begin{align}
i\Sigma&=-\sum_{k_a'}\frac{g^2}{4}\int\frac{d^4k}{(2\pi)^4}\frac{4%
\mathord{\not\mathrel{{\mathrel{k_a'}}}} (\varepsilon^\mu\delta_{p_a+p^{%
\prime}_a,2k_a}+\varepsilon_\psi\Sigma_{45}\delta_{2k_a,p_a-p^{\prime}_a})} {%
((p-k)^{2}-(p_{a}-k_{a})^{2})(k^2-k_a^2)}\nonumber\\
&=\frac{-ig^2}{2(4\pi)^2}\ln\left(\frac{\Lambda^2}{\mu^2}\right) \left[\frac{%
\mathord{\not\mathrel{{\mathrel{p_a}}}}+\mathord{\not\mathrel{{%
\mathrel{p'_a}}}}}{2}(1+\varepsilon_\psi\varepsilon_\psi'e^{(p_a+p'_a)(\pi R)_a/2})\varepsilon^\mu+%
\varepsilon_\psi\frac{\mathord{\not\mathrel{{\mathrel{p'_a}}}}-\mathord{\not%
\mathrel{{\mathrel{p_a}}}}}{2}(1+\varepsilon_\psi\varepsilon_\psi'e^{(p_a-p'_a)(\pi
R)_a/2})\Sigma_{45}\right], \label{ptilde}
\end{align}
where  $\Lambda$ is a cut-off and $\mu$ is renormalization scale.
After transforming to the position space we get
\begin{align}
\delta\mathcal{L}&=\frac{g^2}{8(4\pi)^2}\ln\left(\frac{\Lambda^2}{\mu^2}%
\right)\left[\delta(I)\{\overline{\psi}({i\mathord{\not\mathrel{{%
\mathrel{\partial_a}}}}})(-\varepsilon_\psi \Sigma_{45}
+\varepsilon^\mu)\psi+\overline{\psi}{(i{%
\overleftarrow{\mathord{\not\mathrel{{\mathrel{\partial_a}}}}}})}(-\varepsilon_\psi
\Sigma_{45} -\varepsilon^\mu)\psi\}\right.\nonumber\\
&\quad\quad\quad+\left.\delta(II)\{\overline{\psi}({i\mathord{\not\mathrel{{%
\mathrel{\partial_a}}}}})(-\varepsilon'_\psi \Sigma_{45}
+\varepsilon^\mu)\psi+\overline{\psi}{(i{%
\overleftarrow{\mathord{\not\mathrel{{\mathrel{\partial_a}}}}}})}(-\varepsilon'_\psi
\Sigma_{45} -\varepsilon^\mu)\psi\}\right], \label{fmv}
\end{align}
where
\begin{align}
\delta(I)\equiv \delta(x_a)+ \delta(x_a-\pi R),~~\delta(II)\equiv
\delta(x_a-\pi R/2) +\delta(x_a+\pi R/2),
\end{align}
and $\psi$ is normalized as four dimensional fermion field related
to the six dimensional field by $\psi=\psi^{6D}(2\pi R)$. In our
case the corrections to the self energy of the neutrino will arise
from the diagrams with $W^+_R$, $Z'$, but one can see that these
fields have nonzero mass coming from the breaking of $SU(2)_R$,
thus in Eq.~(\ref{ptilde}) $\tilde p^2 \rightarrow \tilde p^2
+(1-\alpha)M^2_{W^\pm_R,Z'}$. The contribution of the diagram with
$W^\pm_R$ will be
\begin{eqnarray}
\delta
\mathcal{L}=\frac{g_R^2}{8(4\pi)^2}\ln\left(\frac{\Lambda^2}{\mu_{W_R}^2}\right)
\left[\delta(I)\{\overline{\nu_2^R}(-\partial_4-i\partial_5)\nu_2^L+(-\partial_4+i\partial_5)\overline{\nu_2^L}\nu_2^R\}\right.\nonumber\\
+\left.\delta(II)\{\overline{\nu_2^L}(-\partial_4+i\partial_5)\nu_2^R+(-\partial_4-i\partial_5)\overline{\nu_2^R}\nu_2^L\}\right],
\end{eqnarray}
where $\mu_{W_R}^2\sim\mu^2+M^2_{W_R}$. The terms proportional to
the $\delta(I)$ and $\delta(II)$ will lead to the corrections to
the four dimensional action that will have equal magnitude and
opposite sign, so the total correction to the fermion mass will
vanish. The contribution of the diagram with $Z'$ will lead to the
\begin{eqnarray}
\label{deltal} \delta
\mathcal{L}=\frac{g_R^2+g_{B-L}^2}{16(4\pi)^2}\ln\left(\frac{\Lambda^2}{\mu_{Z'}^2}\right)
\left[\delta(I)\{\overline{\nu_2^R}(-\partial_4-i\partial_5)\nu_2^L+(-\partial_4+i\partial_5)\overline{\nu_2^L}\nu_2^R\}\right.\nonumber\\
+\left.\delta(II)\{\overline{\nu_2^L}(\partial_4-i\partial_5)\nu_2^R+(\partial_4+i\partial_5)\overline{\nu_2^R}\nu_2^L\}\right],
\end{eqnarray}
where $\mu_{Z'}^2\sim\mu^2+M^2_{Z'}$. Let us look on the first
term of the formula (\ref{deltal}), it is proportional to the
$\delta(I)\overline{\nu_2^R}\partial_a\nu_2^L$, but one can see
from the KK decomposition (\ref{decompos}), that profiles of the
$\nu_2^R(+,-)$ and $\partial_a\nu_2^L(-,+)$ are both equal to the
$\text{cos}(\frac{n_4x_4+n_5x_5}{R})$ i.e. are maximal at the
$\delta(I)$. The same is true for the others terms of the
(\ref{deltal}), thus the correction to the effective 4D
lagrangian, and KK masses will be \footnote{We are assuming that
at the cut off scale brane like terms are small, and that one loop
brane terms are small compared to the tree level bulk lagrangian,
so to find mass corrections we can use unperturbed KK
decomposition and ignore KK mixing terms , in this approximation
our results coincide with the results presented in
 \cite{aguilla}.}
\begin{eqnarray}
\mathcal{L}_{4D}=\frac{g_R^2+g_{B-L}^2}{4(4\pi)^2}\ln\left(\frac{\Lambda^2}{\mu_{Z'}^2}\right)
\left[\left(\frac{-n_4-in_5}{R}\right)\overline{\nu_2^R}\nu_2^L+\left(\frac{-n_4+in_5}{R}\right)\overline{\nu_2^L}\nu_2^R\right]\nonumber\\
\delta m_{\nu(n_4,n_5)}
=\frac{(g_R^2+g_{B-L}^2)\sqrt{n_4^2+n_5^2}}{4R(4\pi)^2}\ln\left(\frac{\Lambda^2}{\mu_{Z'}^2}\right).
\label{mnu}
\end{eqnarray}
So the correction to the mass of the first KK mode for $\nu_2$
will be:
\begin{eqnarray}
\delta m_{\nu}
=\frac{(g_R^2+g_{B-L}^2)}{4R(4\pi)^2}\ln\left(\frac{\Lambda^2}{\mu_{Z'}^2}\right).
\label{mnu}
\end{eqnarray}

Now we have to evaluate contribution of the diagram Fig.1 (b)
\begin{eqnarray}
i\Sigma =\sum_{k_{a}}f^{2}\int \frac{d^{4}k}{(2\pi )^{4}}\frac{1}{4}\frac{1}{%
(p-k)^{2}-(p_{a}-k_{a})^{2}}\frac{\mathord{\not\mathrel{k}}-%
\mathord{\not\mathrel{k_a'}}}{k^{2}-k_{a}^{2}}[\varepsilon _{\psi
}\Sigma _{45}\delta _{p_{a}-p_{a}^{\prime },2k_{a}}+\varepsilon
_{\phi }\delta _{p_{a}+p_{a}^{\prime },2k_{a}}],
\end{eqnarray}
where $f$ is the 4D Yukawa coupling , and again we will consider
only the terms that are localized at the fixed points of the
orbifold.
\begin{eqnarray}
i\Sigma=\sum_{k_a}f^2\int \frac{d^{4}k}{(2\pi )^{4}}\frac{1}{4}%
\int_0^1d\alpha \frac{(\mathord{\not\mathrel{k}}-\mathord{\not\mathrel{k_a'}}%
)[\varepsilon _{\psi }\Sigma _{45}\delta _{p_{a}-p_{a}^{\prime
},2k_{a}}+\varepsilon _{\phi }\delta _{p_{a}+p_{a}^{\prime },2k_{a}}]}{%
[k^2-k_a^2(1-\alpha)-2(kp)\alpha +p^2\alpha-(p_a-k_a)^2\alpha]^2}
\end{eqnarray}
Proceeding in the same wave as we have done for the diagram with
the vector field we  find
\begin{eqnarray}
i\Sigma =\frac{if^{2}}{16(4\pi
)^{2}}\ln\left(\frac{\Lambda^2}{\mu^2 }\right)
  \left[( \spur{p}-\spur{p_a'}+\spur{p_a})\varepsilon_\psi\Sigma_{45}(1+\varepsilon_\psi\varepsilon'_\psi
  e^{(p_a-p'_a)(\pi R)_a/2})\right.+\nonumber\\
+\left.(\spur{p}-\spur{p_a'}-\spur{p_a})\varepsilon_\phi(1+\varepsilon_\psi\varepsilon'_\psi
e^{(p_a+p'_a)(\pi R)_a/2})
  \right] .
\end{eqnarray}
In our model we have the following Yukawa couplings
\begin{eqnarray}
\overline{\psi}_1^-\Phi\psi_2^+=\overline{\nu}_1\Phi_d^0\nu_2+\overline{e}_1\Phi_d^-\nu_2+\overline{\nu}%
_1\Phi_u^+e_2+\overline{e}_1\Phi_u^0e_2
\end{eqnarray}
So the corrections to the self energy of neutrino will arise from
the diagrams with $\Phi_d^0$ and $\Phi_d^-$. This leads to the
following corrections in the lagrangian
\begin{eqnarray}
\delta {\cal L} =\frac{f^{2}}{8(4\pi
)^{2}}\ln\left(\frac{\Lambda^2
}{\mu^2}\right)[\delta(I)\{\overline{\nu_{2}^R}i
\spur{\partial}\nu_{2}^R +\overline{\nu_{2}^R}
(\partial_4+i\partial_5)\nu_{2}^L+(\partial_4-i\partial_5)\overline{\nu_{2}^L}
\nu_{2}^R\}+\nonumber\\
+\delta(II)\{-\overline{\nu_{2}^L} i\spur{\partial}\nu_{2}^L
+\overline{\nu_{2}^L}
(\partial_4-i\partial_5)\nu_{2}^R+(\partial_4+i\partial_5)\overline{\nu_{2}^R}
\nu_{2}^L\}]
\end{eqnarray}

 The terms proportional to  $\delta(I)$ and $\delta(II)$ lead to
 the corrections to the four dimensional action that will cancel each other, so the
total mass shift due to the diagrams with $\Phi_d^{0,-}$ will be
equal to zero, thus the mass of the neutrino will be corrected
only due to the diagram with the $Z'$ boson (\ref{mnu}).

\section{Corrections to the mass of the gauge boson}
As we have mentioned above the dark matter in the model
\cite{mohapat} is believed to consist from mixture of the KK
photon and right handed neutrinos, so we are interested in the
corrections to the masses of the $B_{Y,a}$ and $B_{Y,\mu}$ bosons.
The lowest KK excitations of the $B_{Y,a}$ and $B_{Y,\mu}$ fields
correspond to  $|p_4|=|p_5|=\frac{1}{R}~$, so everywhere in the
calculations we set ($p_4=p_5\equiv\frac{1}{R})$. At the tree
level both $B_{Y,a}$ and $B_{Y,\mu}$ fields have the same mass
$\frac{\sqrt{2}}{R}$, but radiative corrections can split their
mass levels, and only the lightest one of these two will be stable
and could be the candidate for the dark matter. In this case the
bulk corrections do not vanish by themselves but as was shown in
the \cite{matchev} lead to the same mass corrections for the
$B_{Y,\mu}$ and $B_{Y,a}$ fields.
\begin{figure}[h]
\includegraphics[scale=0.6, angle=90 ]{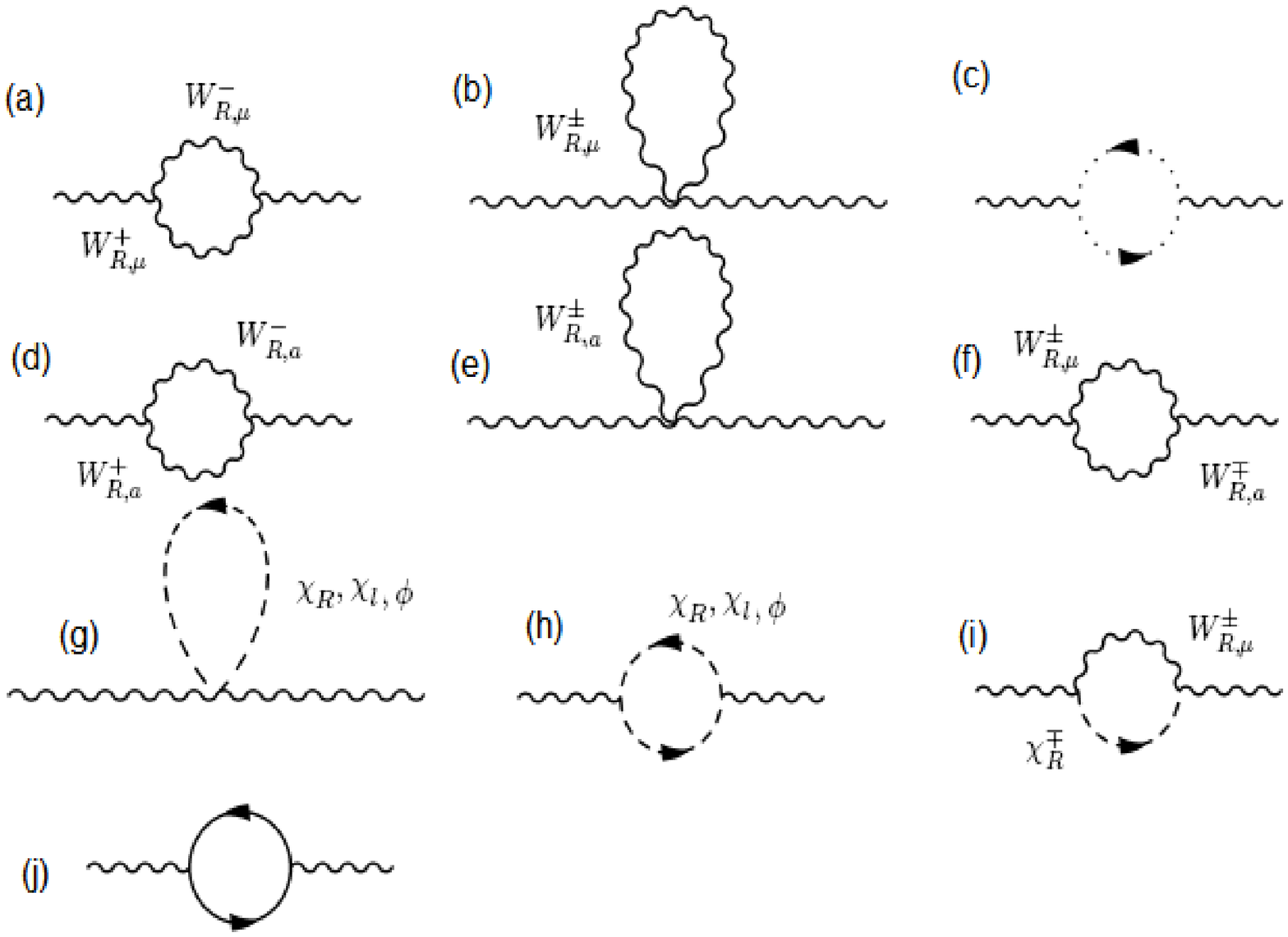} \label{figgf} \caption{self energy diagrams for the
$B_{Y\mu} $ fild:$(a,b)$-loops with $W_{R,mu}^\pm$, $(c)$-ghost
loop, $(d,e,f)$-loops with $W_{R,-}^\pm$, $(g,h,i)$-with goldstone
bosons $\chi_R^\pm$, $(j)$-fermion loop}
\end{figure}

First we will calculate radiative corrections for the $B_{Y,\mu}$
field (calculations are carried out in the Feynman gauge $\xi=1$
), see Fig.2 for the list of the relevant diagrams. The
contribution of
 every diagram can be presented in the form:
\begin{table}
 \caption{Coefficients A,B,C,D for the self energy
diagrams for
 $B_{Y,\mu}$ from the gauge sector and $\chi_R^\pm$}
\begin{tabular}[r]{|c|l|l|l|l|}
\hline
Diagram & A &B & C &D \\
 \hline
$(a)$ & {19}/{3}& -{22}/{3} & 9 &18 \\
\hline
$(b)$&0 & 0 &-6 &-12 \\
\hline
$(c)$&{1}/{3}&{2}/{3} & -1&-2 \\
\hline
$(d)$ & {4}/{3} &-{4}/{3} &-4 &-8\\
\hline
$(e)$&0&0 & 4&8\\
\hline
$(f)$&0&0 & 12&0\\
\hline
$(g)~\chi_R^\pm$&0&0 & -2&-4\\
\hline
$(h)~\chi_R^\pm$& -2/3  &2/3 & 2&4 \\
\hline
$(i)$&0&0 & 0&-4\\
\hline
$(j)$&0&0 & 0&0\\
\hline
\end{tabular}
\end{table}
\begin{eqnarray}
i\Pi_{\mu\nu}=\frac{i}{4(4\pi)^2}\frac{g^2_{B-L}g^2_R}{{g_R^2+g_{B-L}^2}}\ln
\left(\frac{\Lambda^2}{\mu^2_{W_R}}\right)\nonumber\\
\times\left[ A p^2g_{\mu\nu}+Bp_\mu p_\nu  +C
\frac{p_a^2+p_a'^2}{2}g_{\mu\nu}+DM_{W_R}^2g_{\mu\nu}\right]
\end{eqnarray}

The coefficients $A,B,C,D$ are listed in the Table 1. The sum of
all diagrams is  equal to
\begin{eqnarray}
i\Pi_{\mu\nu}=\frac{i}{4(4\pi)^2}\frac{g^2_{B-L}g^2_R}{{g_R^2+g_{B-L}^2}}\ln
\left(\frac{\Lambda^2}{\mu^2_{W_R}}\right) \left[14g_{\mu\nu}
\frac{p_a^2+p_a'^2}{2}+\frac{22}{3}(p^2g_{\mu\nu}-p_\mu p_\nu)
\right].
\end{eqnarray}
this leads to the following corrections to the lagrangian
\begin{eqnarray}
\delta \mathcal{L}=
\left(-\frac{22}{3}(-\frac{1}{4}F_{\mu\nu}F^{\mu\nu})-7
B_{Y\mu}(\partial^2_a B_{Y}^\mu)
\right)\left[\frac{1}{4(4\pi)^2}\frac{g^2_{B-L}g^2_R}{{g_R^2+g_{B-L}^2}}\ln
\left(\frac{\Lambda^2}{\mu^2_{W_R}}\right)\right]\left[\frac{\delta(I)-\delta(II)}{4}\right]
\end{eqnarray}

 These diagrams will not lead to the mass corrections due to the
factor $[\delta(I)-\delta(II)]$. The field $B_Y$  also interacts
with $\chi_L$ and $\phi$, because the $U(1)_Y$ charge is equal to
$Q_{Y}= T^3_R+\frac{Y_{B-L}}{2}$, where $Y_{B-L}$ is $U(1)_{B-L}$
hypercharge. These diagrams will have the same structure as
diagrams (g) and (h), the only difference will be that $\chi_L$
and $\phi$ will have no mass from the breaking of $SU(2)_R$. The
contribution from the fields $\chi_L$ and $\phi_d^{0,+}$ will have
the factor $[\delta(I)-\delta(II)]$, so  only the loops with
$\phi_u^{0,-}$ lead to the nonvanishing result.
\begin{eqnarray}
 \label{Bmu} \delta
\mathcal{L}=\frac{1}{12(4\pi)^2}\frac{g^2_{B-L}g^2_R}{{g_R^2+g_{B-L}^2}}
\ln
\left(\frac{\Lambda^2}{\mu^2}\right)\left[\frac{\delta(I)+\delta(II)}{4}\right]\left(-\frac{1}{4}F_{\mu\nu}F^{\mu\nu}\right)\nonumber\\
\delta
m^2_{B_{Y,\mu}}=-\frac{1}{12(4\pi)^2}\frac{g^2_{B-L}g^2_R}{{g_R^2+g_{B-L}^2}}\frac{2}{R^2}\ln
\left(\frac{\Lambda^2}{\mu^2}\right).
\end{eqnarray}
Now we will calculate the mass corrections for the $B_{Y,a}$
field. At the tree level the mass matrix of the $B_{Y,a}$  arises
from $(-\frac{1}{2}(F_{45})^2)$, and it has two eigenstates:
massless and massive. The massless one is eaten to become the
longitudinal component of the KK excitations of the $B_{Y\mu}$
field, and the massive state behaves like 4D scalar, and is our
candidate for dark matter. In our case ($p_4=p_5=\frac{1}{R}$),
the $(B_+\equiv\frac{B_4+B_5}{\sqrt{2}})$ is the longitudinal
component of the $B_{Y\mu}$, and
$B_-\equiv\frac{B_4-B_5}{\sqrt{2}}$ is the massive scalar.
Nonvanishing mass corrections will arise only from the loops
containing $\phi_u^{0,-}$ fields, the other terms will cancel out
exactly in the same way as for the $B_{Y\mu}$ field.
\begin{eqnarray}
 i\Pi_{ab}=\frac{-i}{4(4\pi)^2}\frac{g^2_{B-L}g^2_R}{{g_R^2+g_{B-L}^2}}[p_c p'_c\delta_{ab}
-p_ap'_b]\ln\left(\frac{\Lambda^2}{\mu^2}\right)\nonumber\\
\label{B_a} \delta
\mathcal{L}=\frac{1}{2}(F_{45})^2\left[\frac{1}{4(4\pi)^2}\frac{g^2_{B-L}g^2_R}{{g_R^2+g_{B-L}^2}}\ln\left(\frac{\Lambda^2}{\mu^2}\right)\right]
 \left[\frac{\delta(I)+\delta(II)}{4}\right]\nonumber\\
\delta
m^2_{B_{Y-}}=-\frac{1}{4(4\pi)^2}\frac{g^2_{B-L}g^2_R}{{g_R^2+g_{B-L}^2}}\frac{2}{R^2}\ln\left(\frac{\Lambda^2}{\mu^2}\right)
\end{eqnarray}
Comparing equations (\ref{B_a}) and (\ref{Bmu}), we see that in
the one loop approximation  $B_{Y-}$ will be the lighter than
$B_{Y\mu}$, so our calculations predict that within the model
\cite{mohapat}, dark matter is admixture of the $B_{Y-}$ and
$\nu_2$ fields. It is interesting to point out that the same
inequality for the radiative corrections to the masses of the
gauge bosons was found in the context of model \cite{ponton}.

\section{Conclusion}

 We studied  the one-loop structure in the field theory in six dimensions compactified on the
$T_2/(Z_2\times Z_2')$ orbifold. We  showed how to take into
account boundary conditions on the $T_2/(Z_2\times Z_2')$ orbifold
and derived  propagators for the fermion, scalar and vector
fields. We calculated mass corrections for the fermion and vector
fields, and then we applied our results to the lightest stable KK
particles in the model \cite{mohapat}. We showed that the lightest
stable modes would be, $B_{Y-}$ and $\nu_2$ fields. These results
are important for the phenomelogical predictions of the model.

\section*{Acknowledgments}
I want to thank R.N.Mohapatra for suggesting the problem and
useful discussion, K.Hsieh for comments. This work was supported
by the NSF grant PHY-0354401 and University of Maryland Center for
Particle and String Theory.

\section*{Appendix }
 In the appendix we will show that contribution of the terms, which do not
 conserve magnitude of the $|p_a|$, will lead to the operators
 localized at the fixed points of the orbifold. We will follow the
 discussion presented in the work of H.Georgi,A.Grant and G.Hailu
 \cite{Georgi} and apply it to our case of $T^2/(Z_2\times Z_2')$
 orbifold. So let us consider general expression.
\begin{eqnarray}
\sum_{p^{\prime}_a=p_a+\frac{m_a}{R}} \frac{1}{2}
\left(1+e^{i\pi(m_4+m_5)}\right)\overline{\psi}(p^{\prime})\Gamma\psi(P)
\end{eqnarray}
where $\Gamma$ is some generic operator, $\psi$ is six-dimensional
fermion field, and factor $1+e^{i\pi(m_4+m_5)}$ appears because
initial and final fields have the same ($Z_2\times Z_2'$)
parities. The action in the momentum space will be given by
\begin{eqnarray}
S=\sum_{p_a}\frac{1}{(2\pi
R)^2}\sum_{p^{\prime}_a=p_a+\frac{m_a}{R}}
\frac{1}{8}\left({1+\varepsilon_e\varepsilon'_e e^{i(\pi
R)_ap_a}}\right)\left(
{1+e^{i\pi(m_4+m_5)}}\right)\left(1+\varepsilon_i\varepsilon'_ie^{i(\pi
R)_a k_a}\right)\overline{\psi} (p^{\prime})\Gamma\psi(p),
\end{eqnarray}
where $\varepsilon_{i,e},~\varepsilon'_{i,e}$ are the $Z_2$ and
$Z_2'$ parities for the particles in the internal and external
lines of the diagram respectively, and $k_a$ is the momentum of
the internal line (we omit integration over the 4D momentum in the
expression). Transforming fields $\psi$ to position space we get
\begin{eqnarray}
S=\frac{1}{8(2\pi R)^2}\sum_{p_a}\sum_{
p^{\prime}_a=p_a+\frac{m_a}{R}}\int dx_a dx^{\prime}_a
e^{-ip^{\prime}_ax^{\prime}_a+ip_ax_a} \cdot
\nonumber\\
\cdot\left[(1+\varepsilon_e\varepsilon_e' e^{i(\pi
R)_ap_a})(1+\varepsilon_i\varepsilon'_i e^{i(\pi R)_a(p_a\pm
p_a\pm m_a/R )/2})
(1+e^{i\pi(m_4+m_5)})\right]\overline{\psi}(x^{\prime})\Gamma\psi(x),
\end{eqnarray}
the upper and lower signs in the expression
$(1+\varepsilon_i\varepsilon_i' e^{i(\pi R)_a(p_a\pm p_a\pm m_a
/R)/2})$ correspond to the $k_a=\frac{p_a\pm p'_a}{2}$ in the
propagator. Now we can use identities:
\begin{eqnarray}
\sum_{p_a} \frac{e^{ip_a(x_a-x^{\prime}_a)}}{(2\pi R)^2}=\delta(x_a-x^{%
\prime}_a)  ,\nonumber \\
\sum_{m=-\infty}^\infty e^{\frac{imx}{R}}=\sum_{m=-\infty}^{\infty}\delta(m-%
\frac{x}{2\pi R}).
\end{eqnarray}
so
\begin{align}
S=\frac{1}{4}\int d^{(6)}x~ \overline{\psi}(x)\Gamma\psi(x)
\sum_{m_a} \left[\delta(m_a-\frac{x_a}{2\pi R})+
\delta(m_a-\frac{1}{2}-\frac{x_a}{2\pi R })\right. +\nonumber\\
+\left.\left
(\delta(m_a-\frac{1}{4}-\frac{x_a}{2\pi R})
+\delta(m_a+\frac{1}{4}-\frac{x_a}{2\pi R }) \right) \cdot
\begin{cases}
(\varepsilon_i\varepsilon_i')(\varepsilon_e\varepsilon_e')~
\text{for } k_a=\frac{p_a+ p'_a}{2} \\
\varepsilon_i\varepsilon_i'~ \text{for } k_a=\frac{p_a-
p'_a}{2}\end{cases} \right] \label{local}
\end{align}
So the brane terms will be localized at the  points $(0,0),(\pi
R,\pi R), (\pm\pi R/2,\pm\pi R/2)$.

\end{document}